\documentclass[a4paper]{jpconf}
\usepackage{graphicx}
\begin{document}
\title{Quantum scattering beyond the plane-wave approximation}

\author{Dmitry Karlovets}

\address{Tomsk State University, Lenina 36, 634050 Tomsk, Russia}


\begin{abstract}
While a plane-wave approximation in high-energy physics works well in a majority of practical cases,
it becomes inapplicable for scattering of the vortex particles carrying orbital angular momentum, of Airy beams, of the so-called Schr\"odinger cat states, 
and their generalizations. Such quantum states of photons, electrons and neutrons have been generated experimentally in recent years, 
opening up new perspectives in quantum optics, electron microscopy, particle physics, and so forth. 
Here we discuss the non-plane-wave effects in scattering brought about by the novel quantum numbers of these wave packets. For the well-focused electrons of intermediate energies, 
already available at electron microscopes, the corresponding contribution can surpass that of the radiative corrections. 
Moreover, collisions of the cat-like superpositions of such focused beams with atoms allow one to probe effects of the quantum interference, 
which have never played any role in particle scattering.
\end{abstract}

\section{Introduction}

When studying scattering of particles, one usually treats their states as the eigenstates of a momentum operator, 
that is, as the plane waves, $\hat{\bf p}|{\bf p}\rangle = {\bf p}|{\bf p}\rangle$. The only objection one can raise is that the plane waves, being completely delocalized in space and time, do not exist in Nature. 
That is why the well-normalized wave packets are oftentimes considered first and a limit of the infinitely narrow ones in momentum space is taken when passing to the cross section. 
And yet the plane-wave approximation successfully describes the overwhelming majority of the processes studied at modern colliders. 
On the other hand, a proper discussion of its limits of applicability in the scattering problems seems to be absent in literature,
which is a bit disturbing, especially when striving for higher-order radiative corrections. There are also important examples from the collider physics, atomic physics, neutrino physics, etc., 
in which the plane-wave model becomes inapplicable -- see, for instance, \cite{Impact, Akhmedov_10, t1, I_S, JHEP, PRL} and references therein.


In mathematical physics, a lot of non-plane-wave solutions for the Dirac equation are well known \cite{Bagrov_Mono}.
Yet it was only in the 1990s that the first photons possessing a well-defined orbital angular momentum (OAM) were generated \cite{Mono}. 
Such a twisted photon, $|\omega, k_z, \kappa, \ell, \lambda\rangle$, is a (paraxial) monochromatic state with the OAM $\ell$, a helicity $\lambda$, 
a longitudinal momentum $k_z$, and an absolute value of the transverse momentum $\kappa$. During the last several years, 200-300-keV electrons and cold neutrons carrying the OAM 
were also generated \cite{Uchida, Verbeeck, McMorran, neu}, along with the so-called Airy photons and electrons \cite{Airy_Exp, Airy_El_Exp} 
and their different generalizations. Possible applications of these states to the hadronic physics have been discussed in Refs.\cite{JHEP, Igor_phase, Dodonov}.

Here we study how scattering processes with these particles can be described within an S-matrix formalism. 
A non-plane-wave contribution to the cross section turns out to be attenuated by a ratio of the Compton wave length, $\lambda_c = \hbar/(mc)$, 
to the beam width $\sigma_x$, $\lambda_c/\sigma_x$, which can reach the values of $\lambda_c/\sigma_x \sim 10^{-3}$ for electrons, 
but stays several orders of magnitude smaller for available hadronic beams. Thus for the well-focused electrons of intermediate energies, the corresponding contribution 
can even surpass that of the radiative corrections, as the former is attenuated as $f (s,t)\, \lambda_c/\sigma_x$, being no longer negligible when $f (s,t)$, 
which depends on a phase $\zeta_{fi}$ of a scattering amplitude $M_{fi}$, is not vanishing. Moreover, colliding coherent superpositions of the wave packets with atoms, one can get out of the paraxial approximation when the beams are focused to an atomic size, which is already feasible. 
The units with $\hbar = c = 1$ are used.

\section{Qualitative considerations}

Let us consider a process of $2\rightarrow N$ scattering. 
Staying first within the common plane-wave model, the quantum effects arise because of recoil, because of such intrinsic degrees of freedom as, for instance, spin, and because of the radiative corrections. 
The latter are oftentimes said to be the ``genuinely quantum'' phenomena, and in a perturbative regime their contribution 
is attenuated by $\alpha \ll 1$ (in QED, $\alpha= 1/137$). Going beyond the plane-wave approximation, however, we get yet another source of the genuinely quantum corrections, 
which contribute already to the tree level. These are evoked by the wave nature of particles, with their de Broglie wave length, $\lambda_{dB} = 2\pi/p$,
being a characteristic parameter. The plane-wave approximation is usually said to be applicable when the wave packets describing particles are narrow in the momentum space, that is, 
\begin{eqnarray}
& \displaystyle \frac{\sigma_p}{|\langle{\bf p}\rangle|} \sim \frac{\lambda_{dB}}{\sigma_x} \ll 1,
\label{dB_1}
\end{eqnarray}
where $\sigma_p \sim 1/\sigma_x$ is a packet's width and $\langle{\bf p}\rangle$ is its mean momentum.
This inequality is manifestly not Lorentz invariant. 
Let $\sigma_x$ be a packet's width in a plane perpendicular to the direction of a boost, which is made along a collision axis for definiteness. 
Then a Lorentz-invariant generalization of (\ref{dB_1}) would be
\begin{eqnarray}
& \displaystyle \frac{\sigma_p}{m} \sim \frac{\lambda_c}{\sigma_x} \ll 1,
\label{dB_2}
\end{eqnarray}
with $\lambda_c$ instead of $\lambda_{dB}$. Note that in a more general case of arbitrary boost, one should deal with a tensor $\sigma_{ij}$ instead of the scalar $\sigma_p$. 
As a result, the small parameter of the problem no longer preserves its simple form (\ref{dB_2}) \cite{JHEP}.

For ultrarelativistic energies, $|\langle{\bf p}\rangle| \gg m$, the de Broglie wave length is much smaller than the Compton one, which does not mean, however, that the wave nature of particles does not reveal itself in scattering. As we demonstrate below, in fact it does 
and the non-plane-wave contribution to the (Lorentz invariant) cross section is attenuated by the parameter (\ref{dB_2}) 
and it does not decrease with the increase of the energy. The smallness of (\ref{dB_2}) also necessitates the applicability of a one-particle theory with no negative-energy solutions.

\section{Corrections to the plane-wave cross section}

The plane-wave cross section $d\sigma^{(pw)}$ depends on an absolute value of the amplitude $|M_{fi} ({\bf p}_i)|^2$ with ${\bf p}_i$ being the incoming momenta. 
When dealing with the wave packets instead, one gets 
\begin{eqnarray}
& \displaystyle|M_{fi}({\bf p}_i)|^2 \rightarrow M_{fi}({\bf p}_i) M_{fi}^*({\bf p}_i^{\prime})\ or\ M_{fi}({\bf p}_i + {\bf k}_i/2) M_{fi}^*({\bf p}_i- {\bf k}_i/2)
\label{M}
\end{eqnarray}
where ${\bf p}_i$ and ${\bf p}_i^{\prime}$ share the same mean value $\langle{\bf p}_i\rangle$, but the mean value of ${\bf k}_i$ is vanishing. 
Then, presenting the complex amplitude as $M_{fi} = |M_{fi}|\, \exp\{i\zeta_{fi}\}$, we obtain
\begin{eqnarray}
& \displaystyle M_{fi}({\bf p}_i + {\bf k}_i/2) M_{fi}^*({\bf p}_i- {\bf k}_i/2) = \left (|M_{fi}({\bf p}_i)|^2 + \mathcal O (k_i^2)\right )\exp\left\{i{\bf k}_i \frac{\partial\zeta_{fi}}{\partial{\bf p}_i} + \mathcal O (k_i^3)\right\},
\label{M_2}
\end{eqnarray}
where the neglected terms are negligible when Ineq.(\ref{dB_2}) holds true, akin to the paraxial or WKB approximation. 
Hence, the first correction to the plane-wave cross section is linearly attenuated by $\lambda_c/\sigma_x$ and depends on derivatives of the phase $\zeta_{fi}$,
\begin{eqnarray}
& \displaystyle \frac{d\sigma^{(1)}}{d\sigma^{(pw)}} = \frac{\lambda_c}{\sigma_x} \left (f_1 (s,t) \frac{\partial\zeta_{fi}}{\partial s}+ f_2 (s,t)\frac{\partial\zeta_{fi}}{\partial t}\right ),
\label{dsigma1}
\end{eqnarray}
with $s,t$ being the Mandelstam variables. It is these functions, $f_1$ and $f_2$, that depend on such quantum numbers of the incoming particles as, for instance, the OAM or the Airy phase.

One can infer from (\ref{M_2}) that these functions do not vanish only when the in-state can be characterized with a polar vector 
${\bf b}_i$: ${\bf k}_i \rightarrow {\bf b}_i$. In other words, the initial system of particles as a whole must possess a finite \textit{dipole moment}. 
The simplest way to realize this is to collide two beams head on, but at a finite impact parameter. An alternative would be to collide two non-symmetric wave packets like Airy beams 
or vortex electrons with a center of the ring slightly shifted from that of the beam. The latter vortex beams can be shown to possess a finite uncertainty of the OAM.
This finite uncertainty plays a crucial role, as the pure Bessel states are orthogonal and, hence, neither overlap nor interfere, exactly like the plane waves. 
It is a finite overlap of the incoming packets that results in a non-vanishing $d\sigma^{(1)}$ from (\ref{dsigma1}), and that is why the mutual disposition of the packets (in either $x$- or $p$-space) 
comes into play \cite{JHEP}. 


Next, such a dipole moment is obviously a potential source of \textit{an azimuthal asymmetry} in scattering, as the functions $f_1, f_2$ 
can now depend on the products ${\bf b}\cdot {\bf p}_f$ with ${\bf p}_f$ being a particle's final momentum. 
The asymmetry can reveal itself, say, in $ee$, $e^+e^-$, $pp$, or $p\bar{p}$ collisions (see also \cite{Igor_phase, Dodonov}), 
and it can be measured, for instance, by taking two electron beams, 
each focused to a spot of $0.1$ nm, as in the Refs.\cite{Angstrom, Pohl}, and colliding them slightly off-center at an impact parameter of $b \sim \sigma_x$. 
The order of the predicted asymmetry, very roughly, is
\begin{eqnarray}
& \displaystyle \frac{d\sigma^{(1)}}{d\sigma^{(pw)}} \sim \frac{\lambda_c}{\sigma_x} \sim 10^{-4} - 10^{-3}.
\label{A}
\end{eqnarray}
The more detailed calculations support this estimate \cite{JHEP}. 

\section{Going beyond the WKB regime}

The paraxial approximation fails to work when $\sigma_x \sim \lambda_c$, and the one-particle theory itself becomes inapplicable. 
In order to understand better what effects are lost when (\ref{dB_2}) holds true, let us recall that the RHS of (\ref{M}) represents just a density matrix of the process,
\begin{eqnarray}
& \displaystyle M_{fi}({\bf p}_i + {\bf k}_i/2) M_{fi}^*({\bf p}_i- {\bf k}_i/2) \rightarrow \rho_{fi} ({\bf p}_i, {\bf k}_i),
\label{Mrho}
\end{eqnarray}
a Fourier transform of which is called a Wigner function,
\begin{eqnarray}
& \displaystyle n_{fi}({\bf r}_i, {\bf p}_i) = \int \frac{d^3 k}{(2\pi)^3}\,\, e^{i{\bf k}_i{\bf r}_i}\rho_{fi} ({\bf p}_i, {\bf k}_i),
\label{MW}
\end{eqnarray}
studied, for instance, in \cite{Wigner}. Clearly the expansion (\ref{M_2}) leaves this Wigner function \textit{positive everywhere}, that is, quasi-classical.

The same holds for the incoming wave packets. If a wave function of an in-state is $\psi ({\bf p}_i)$, 
a square of the matrix element (\ref{Mrho}) depends on $\psi({\bf p}_i + {\bf k}_i/2) \psi^*({\bf p}_i- {\bf k}_i/2) \rightarrow \rho ({\bf p}_i, {\bf k}_i)$
with $\rho$ being a particle's density matrix. In the perturbative regime (\ref{dB_2}), an expansion analogous to (\ref{M_2}) would also leave the corresponding Wigner functions 
positive everywhere. It has a clear interpretation: the quasi-classical states with the non-negative Wigner functions are nearly coherent, 
whereas those with the not-everywhere positive functions experience \textit{destructive self-interference}. 
For a single packet, the negative regions become noticeable only when it is focused to a spot comparable to $\lambda_c$. 
From this one may deduce that it is practically impossible to get out of the WKB regime.

There are, however, possibilities to go beyond the paraxial approximation even without such a tight focusing.
Indeed, the non-classical states of light, such as a so-called Schr\"odinger cat state, 
are known to reveal a high degree of non-classicality even if they are not focused below a shot-noise limit. 
Taking instead of a single (Gaussian) packet a quantum superposition of them, which may not be Gaussian because of the quantum interference, 
we get a cat-like system with a Wigner function whose negative regions become noticeable at a much larger scale than $\lambda_c$. 

The next step is to collide such nonclassical states with the composite ``particles'' like atoms or molecules 
having an effective size of $a \gg \lambda_c$. Then the parameter (\ref{dB_2}) is to be changed to $a/\sigma_x$,
which can be of the order of unity, even if (\ref{dB_2}) still holds. For a hydrogen atom, $a$ is just a Bohr radius, $a \approx 0.053$ nm. 
As beams of the electrons with $\sigma_x \sim 0.1$ nm have been already obtained \cite{Angstrom, Pohl}, one can probe the highly nonclassical regime of scattering, 
in which negative values of the Wigner functions give noticeable contribution to the cross section. 
The simplest effect here would also be an azimuthal asymmetry in elastic scattering of a Schr\"odinger cat state off an atom,
which would be of the order of $a/\sigma_x \sim 10^{-1}$ (see details in \cite{PRL}).


\section{Conclusion}
The non-plane-wave effects in quantum scattering are no longer vanishing for the focused beams of electron microscopes, 
which allows one to probe a contribution of the Coulomb phase of the scattering amplitude to the cross section. 
These effects can even surpass those of the radiative corrections, and they become only moderately attenuated beyond the paraxial approximation, 
which is already feasible with current technologies. For hadronic beams, however, the corresponding contribution seems to be several orders of magnitude smaller. 
On the other hand, quantum interference can also play a role in describing final jets in hadronic collisions \cite{Dodonov},
which is yet to be studied in more detail.

\ack
I am grateful to V.\,G.~Bagrov, I.\,F.~Ginzburg, I.\,P.~Ivanov, P.\,O.~Kazinski, P.~Kruit, A. Di Piazza, V.\,G.~Serbo, O.\,V.~Teryaev, J.~Verbeeck, and A.\,S.~Zhevlakov for useful discussions and criticism.
This work is supported by the Russian Science Foundation (project No.\,17-72-20013).
 
\

\end{document}